\documentclass[prl,amssymb,superscriptaddress, longbibliography,twocolumn]{revtex4-2}
\usepackage{amsmath}
\usepackage{amssymb}
\usepackage{braket}
\usepackage{hyperref}
\usepackage[normalem]{ulem}
\usepackage{mathtools}
\usepackage{mathrsfs}
\usepackage{float}
\usepackage[export]{adjustbox}

\newcommand{\be}{\begin{equation}}
\newcommand{\ee}{\end{equation}}

\newcommand{\ra}{\rangle}
\newcommand{\la}{\langle}
\renewcommand{\a}{\alpha}
\newcommand{\bea}{\begin{eqnarray}}
\newcommand{\eea}{\end{eqnarray}}

\renewcommand{\a}{\alpha}

\usepackage{xcolor}
\usepackage{color}

\usepackage{hyperref}
\hypersetup{colorlinks=true,citecolor={blue},linkcolor={blue},urlcolor={blue}}

\newcommand\alexout{\bgroup\markoverwith{\textcolor{blue}{\rule[0.5ex]{1pt}{1pt}}}\ULon}

\begin{document}

\preprint{APS/123-QED}

\title{Nonlinearity-driven Topology {\sl via} Spontaneous Symmetry Breaking}

\author{Alessandro Coppo\textsuperscript{\textcolor{blue}{*}}\hspace*{-0.5mm}}
\affiliation{Istituto dei Sistemi Complessi, Consiglio Nazionale delle Ricerche, Via dei Taurini, 19 I-00185 Roma (IT)}
\affiliation{Physics Department,  University of Rome, ``La Sapienza'', P.le A. Moro, 2 I-00185 Roma (IT)}
\author{Alexandre Le Boit\'e}
\affiliation{Universit\'e Paris Cit\'e, CNRS, Mat\'eriaux et Ph\'enom\`enes Quantiques, F-75013 Paris, France}
\author{Simone Felicetti}%
\affiliation{Istituto dei Sistemi Complessi, Consiglio Nazionale delle Ricerche, Via dei Taurini, 19 I-00185 Roma (IT)}
\affiliation{Physics Department,  University of Rome, ``La Sapienza'', P.le A. Moro, 2 I-00185 Roma (IT)}
\author{Valentina Brosco}%
\affiliation{Istituto dei Sistemi Complessi, Consiglio Nazionale delle Ricerche, Via dei Taurini, 19 I-00185 Roma (IT)}
\affiliation{Physics Department,  University of Rome, ``La Sapienza'', P.le A. Moro, 2 I-00185 Roma (IT)}

\begin{abstract}
\vspace*{-4mm}
\begin{center}\textcolor{blue}{$^*$}(\href{mailto:alessandro.coppo@cnr.it}{\footnotesize alessandro.coppo@cnr.it})\end{center}
\vspace*{2mm}
Topology and nonlinearity are deeply connected. However, whether topological effects can arise solely from the structure of nonlinear interaction terms, and the nature of the resulting topological phases, remain to large extent open questions. Here we consider a chain of  parametrically-driven quantum resonators coupled only via weak nearest-neighbour cross-Kerr interaction, without any quadratic tunneling term. We show that, when the drive overcomes a critical threshold value,  the system undergoes a transition from the atomic limit of decoupled oscillators to a symmetry-broken topological phase. The topology is dictated by the structure of the Kerr nonlinearity,  yielding a non-trivial bulk-boundary correspondence. In the topological phase, we find different effective models for periodic and open boundary conditions and derive analytical approximations  for the low-energy spectrum, identifying the conditions to observe topological edge modes. 
\end{abstract}

\maketitle

\noindent In classical and quantum many-body systems, phase transitions between states with different symmetries are described by local order parameters. The scaling properties of these order parameters near the transition allow for the identification of universal features and the classification of critical behavior \cite{sachdev2011}. Unlike symmetry-breaking phases, topological phases are defined not by local order parameters but by global topological invariants \cite{kitaev2009, hasan2010, chiu2016}. In the absence of nonlinearities, topological band theory provides a unifying framework for characterizing topological phases in quantum materials \cite{kane2013} and photonic systems \cite{ozawa2019} enabling the description of topological phenomena such as quantized transport \cite{thouless1982, thouless1983} and gapless edge states \cite{jackiw1976, hatsugai1993}.

Going beyond quadratic Hamiltonians, non-linear interactions can profoundly influence topological systems \cite{rachel2018}. For example, interactions can modify or destroy existing topological effects \cite{juergensen2023,mostaan2022,sone2024}, perturb boundary modes \cite{jezequel2022, isobe2024,brunelli2023}, or give rise to topologically ordered phases \cite{wen1990,kitaev2006, levin2006,su1981, arovas1984, wen1995}. These phenomena demonstrate the deep connection between topology and nonlinearity. However, whether topological effects can arise solely from the structure of nonlinear interaction terms, and the nature of the resulting topological phases, remain to large extent  open questions. Beside its fundamental relevance, answering these question is crucial to properly characterize topological effects and discover novel topological materials.

Quantum resonators  have proved to be compelling systems for exploring both topological properties~\cite{lu_topological_2014} and nonlinear quantum phenomena~\cite{carusotto_quantum_2013, gu_microwave_2017}. One of the striking manifestations of nonlinearity has been the prediction~\cite{carmichael_breakdown_2015,casteels_power_2016,bartolo2016,dykman2012,minganti_non-gaussian_2023} and observation~\cite{fitzpatrick_observation_2017,brookes_critical_2021,chen_quantum_2022,zheng_observation_2023,beaulieu_observation_2023} of driven-dissipative quantum phase transitions.   
Even in finite-component systems at equilibrium, quantum phase transitions can formally emerge when the quadratic part of the effective resonator potential becomes unstable~\cite{ashhab_superradiance_2013,hwang_quantum_2015,peng_unified_2019}. Nonlinearities, even vanishingly small, are then essential in establishing a stable phase~\cite{felicetti_universal_2020}.  Upon crossing the critical point, the system transitions from a low-excitation phase, to a symmetry-broken highly-populated state.
 
Parametric quantum resonators are a class of systems where such instabilities occur and  typically
originate from two-photon generation processes.
They can be realized in a variety of settings \cite{dykman2012} ranging form nanomechanical systems \cite{dykman2012,poot2012} to optical oscillators \cite{drummond1980,arenz2013,drummond1981,calvanesestrinati2024} and superconducting circuits~\cite{beaulieu_observation_2023}.
Their application range from the encoding of Schr\"odinger cat  qubits~\cite{puri2017, goto2019,grimm2020}, to quantum sensing~\cite{dicandia2023,beaulieu_criticality-enhanced_2024,petrovnin_microwave_2023,gu_quantum_2024} and large-scale simulations~\cite{kanao2021}.
In all these frameworks, systems of multiple parametric quantum resonators  have been shown to enable novel interesting functions 
including  dissipative bath engineering~\cite{zapletal2022}, hyperspin encoding~\cite{calvanesestrinati2022}  and collective quantum sensing~\cite{alushi2024} as well as the simulation of  non-equilibrium quantum phases~\cite{caleffi2023}.
Several models have been considered in recent years to explore topological properties in arrays of bosonic resonators, including the addition of parametric processes~\cite{roy2022, guo2023}, counter-rotating terms~\cite{goren_topological_2018} or local nonlinear terms~\cite{Leykam2016, hadad2016,sone2025,Bloch25} to  topological tight-binding Hamiltonians, or the breaking of time-reversal symmetry with external fields~\cite{bardyn2016, rassaert2024}. 

In this letter, we adopt an alternative approach and investigate a setting in which effective topological bands emerge from Spontaneous Symmetry Breaking (SSB).  We consider a chain of parametric quantum resonators featuring  weak nonlinearities-- local Kerr and staggered cross-Kerr terms-- without any direct  quadratic coupling between the resonators. We focus on the properties of the system above threshold, where the quadratic potential is unstable. We show that nonlinear interaction terms can induce a transition from the \emph{atomic limit} of decoupled oscillators to a non-trivial \emph{topological phase}. We analyze both the mean-field phase-diagram, and the nature of the emerging bosonic oscillations around equilibrium solutions. The phase diagram features different phases including a density-wave phase characterized by localized Gaussian excitations, and a homogeneous phase with dispersive Gaussian modes described by an effective Su-Schrieffer-Heeger (SSH) Hamiltonian \cite{su1979}.
Interestingly, the structure of this Hamiltonian depends on the boundary conditions, yielding a breakdown of the bulk-boundary correspondence, such that a nonzero Zak phase \cite{zak1989} does not guarantee the emergence of protected Gaussian edge modes. Finally, we show how topological edge modes can be restored by small corrections to the boundary sites.\\ 
\begin{figure}
\hspace*{-5mm}\includegraphics[scale=0.43]{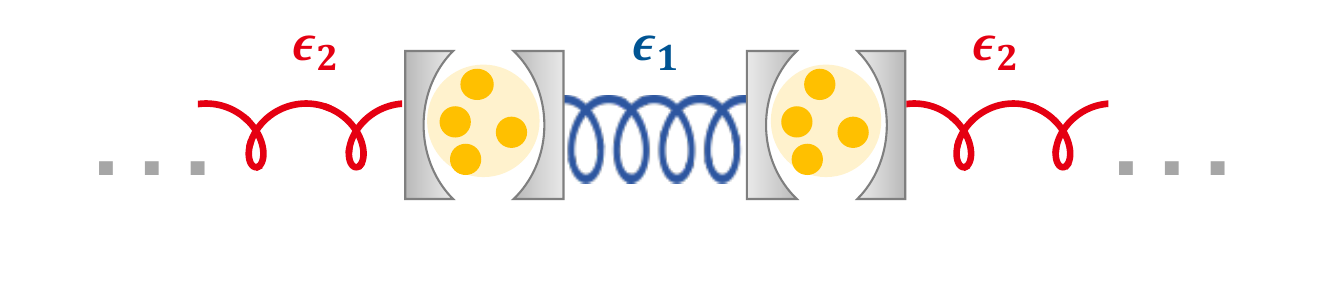}
   \caption{\textbf{Model.} Sketch of the parametric quantum resonator chain with staggered cross-Kerr interactions, see Eqs. (\ref{e.H0}-\ref{e. Hamiltonian cross-Kerr}) }
    \label{f.Kerr chain}
\end{figure}

\textit{Model -- }  We consider  a chain of $2N$ Kerr resonators with frequency $\omega$ and  parametric driving $\lambda$, as sketched in Fig.~\ref{f.Kerr chain}.  We assume that nearest-neighbor resonators are coupled only {\sl via} cross-Kerr interaction terms. Introducing bosonic creation/annihilation operators, the system Hamiltonian can thus be written as the sum of two terms as $H=H_L+H_C$ where $H_L$ describes a chain of decoupled parametric Kerr resonators ($\hbar = 1$)
\be H_L\!=\!\!\!\!\!\!\!\! \sum_{\substack{1\leq n\leq N \\ I\in\{A,B\}}}\!\!\!\!\big[\omega  c^\dag_{In}c_{I n}+\frac{\lambda}{2}\left( c^\dag_{In}c^\dag_{I n} +{\rm H.c.}\right)+ \epsilon_L\, c^\dag_{In}c^\dag_{I n} c_{In}c_{I n}\big]\label{e.H0}\ee
while $H_{C}$ accounts for cross-Kerr nonlinearities,
\be
 H_{C}=\!\sum_{n = 1}^N\big[\epsilon_1\,c^\dag_{An}c_{A n} c^\dag_{Bn}c_{B n}+ \epsilon_2\, c^\dag_{An+1}c_{A n+1} c^\dag_{Bn}c_{B n} \big]
   \label{e. Hamiltonian cross-Kerr}
\ee
where we note that the local Kerr intensity,  $\epsilon_L$, is homogeneous while the cross-Kerr nonlinearity has a staggered structure. 
\begin{figure}
\centering
  \hspace*{-5mm}\includegraphics[scale=0.55]{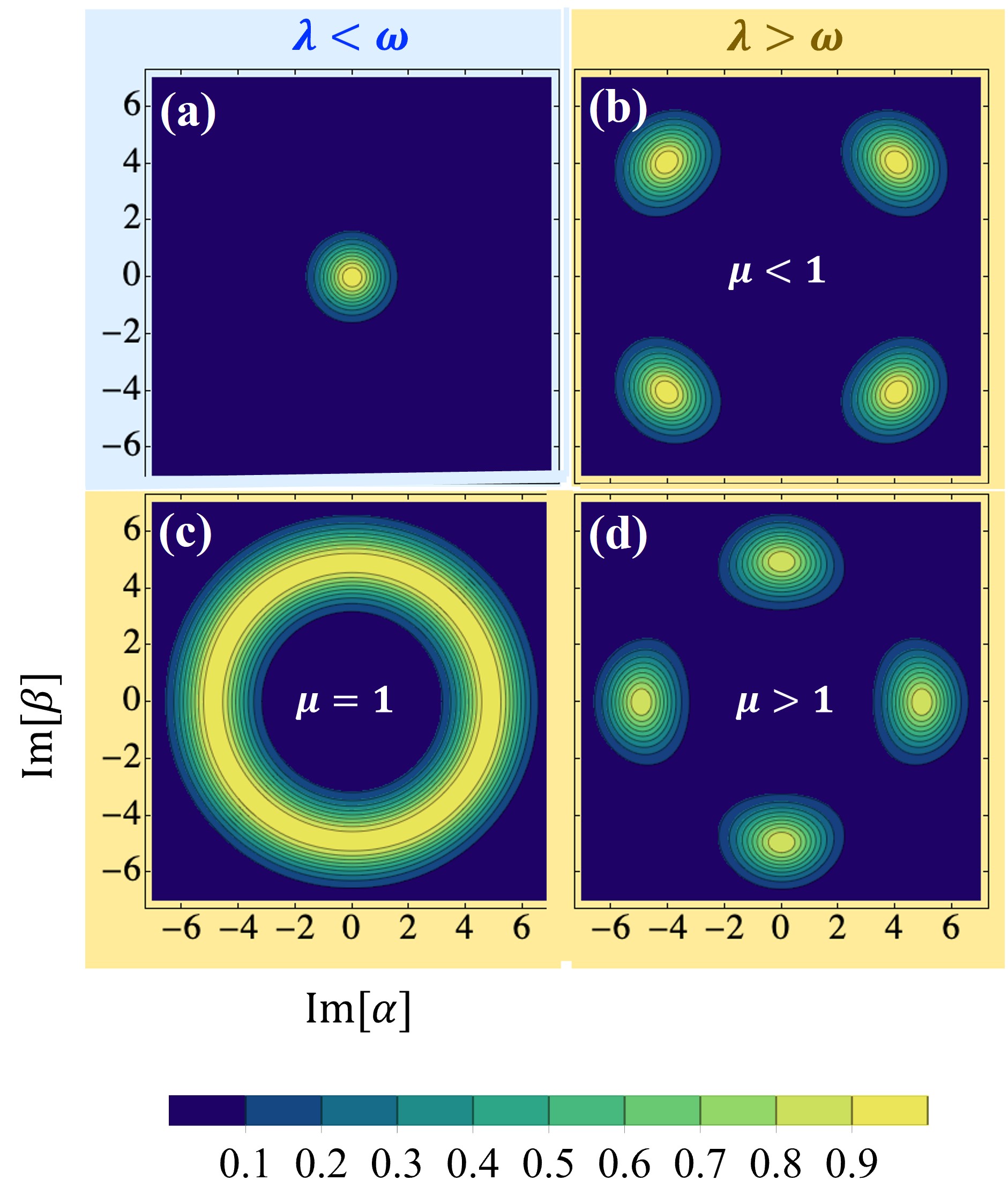}
   \caption{\textbf{Single-cell Husimi function.} Structure of the ground state Husimi function of two resonators coupled via the cross-Kerr nonlinearity $\epsilon_1$. Parameters values: (a) $\lambda=0$ and $\epsilon_1=\epsilon_L$; (b-d) $\lambda=2\omega$ and, $\epsilon_1=\epsilon_L$ (b), $\epsilon_1=2\epsilon_L$ (c) and $\epsilon_1=3\epsilon_L$ (d). In all panels we set $\epsilon_L=0.02~\omega$, we consider $50$ Fock states for each resonator and we renormalize the function to its maximum.}
    \label{f.two_mode_Husimi}
\end{figure}
Starting from the Hamiltonian $H$ we can define  three adimensional parameters, respectively indicated as $g$, $\mu$ and $\delta$, that govern the physics of the model.
The parameter $g$ sets the distance from the instability for each resonator and it reads
\be
g^2=\frac{\lambda-\omega}{2\epsilon_L}
\ee
with $\lambda>\omega$. The parameters $\mu$ and $\delta$ define the structure of the nonlinearity and they are given by
\be
\mu=\frac{\epsilon_2+\epsilon_1}{ 2\epsilon_L}\quad {\rm and} \quad \delta=\frac{\epsilon_2-\epsilon_1}{\epsilon_2+\epsilon_1}.
\ee\\
\\
\textit{Single-cell --}  To understand how the topological symmetry-broken phase emerges, it is useful to start by setting the inter-cell cross Kerr to zero
$\epsilon_2=0$ and considering a single resonator pair.
The ground-state $\Ket{G}$ can be obtained by exact numerical simulations. We plot in Fig.~\ref{f.two_mode_Husimi} the squared modulus of the Husimi function, $Q(\alpha,\beta)=\braket{\alpha,\beta|G}$, for different values of the ratios $\lambda/\omega$ and $\mu$, with $\alpha,\beta\in\mathbb{C}$  identifying the coherent states of the two resonators.
 
For $\lambda\ll\omega$ the ground state is essentially the vacuum and $Q(\alpha,\beta)$ is a  Gaussian centered in $\alpha=\beta=0$ as shown in Fig.~\ref{f.two_mode_Husimi}(a). As $\lambda$ increases, the ground state becomes more and more squeezed. However, as long as $\lambda < \omega$, the function $Q(\alpha, \beta)$ remains localized around $\alpha = \beta = 0$.  
 When $\lambda$ overcomes the threshold $\omega$,   the vacuum state becomes unstable and three distinct regimes emerge, as illustrated in Fig.~\ref{f.two_mode_Husimi}(b-d). 
 
When cross-Kerr interaction dominates over the local one, {\sl i.e.} for $\mu>1$, $Q(\alpha,\beta)$ has of four maxima located at $\{\pm i g,0\}$ and $\{0,\pm i g\}$, 
as shown in Fig.\ref{f.two_mode_Husimi}(d). In this case, it is energetically favorable to frustrate cross-Kerr interaction by creating an inhomogeneous semiclassical ground-state, consistent with the findings of Ref.~\cite{hellbach2024}. In this regime, each lobe of the Husimi function represents a single-mode squeezed state, and the corresponding quadratic Hamiltonian does not couple modes in the two resonators.  

When the local Kerr interaction dominates, for $\mu < 1$, both resonators are equally excited and the maxima of $Q(\alpha, \beta)$ shift to $\{\pm i \bar{g}, \pm i \bar{g}\}$, where $\bar{g}^2 = g^2/(\mu + 1)$, as depicted in Fig.~\ref{f.two_mode_Husimi}(b). In this regime, each lobe of the Husimi function represents a two-mode squeezed state, indicating an effective coupling between the two resonators in the quadratic Hamiltonian.
Eventually, for  $\mu=1$,  the maxima of the Husimi function lie on a circle of radius $\bar{g}$ as shown in  Fig.~\ref{f.two_mode_Husimi}(c). 
Fig.~\ref{f.two_mode_Husimi} therefore illustrates how, for  $\epsilon_2=0$, the structure of the ground state, the squeezing properties, and consequently the nature of the Bogoliubov excitations change across the three regimes.\\

\begin{figure}[H]
\centering
\vspace*{5mm}
    \includegraphics[scale=0.55]{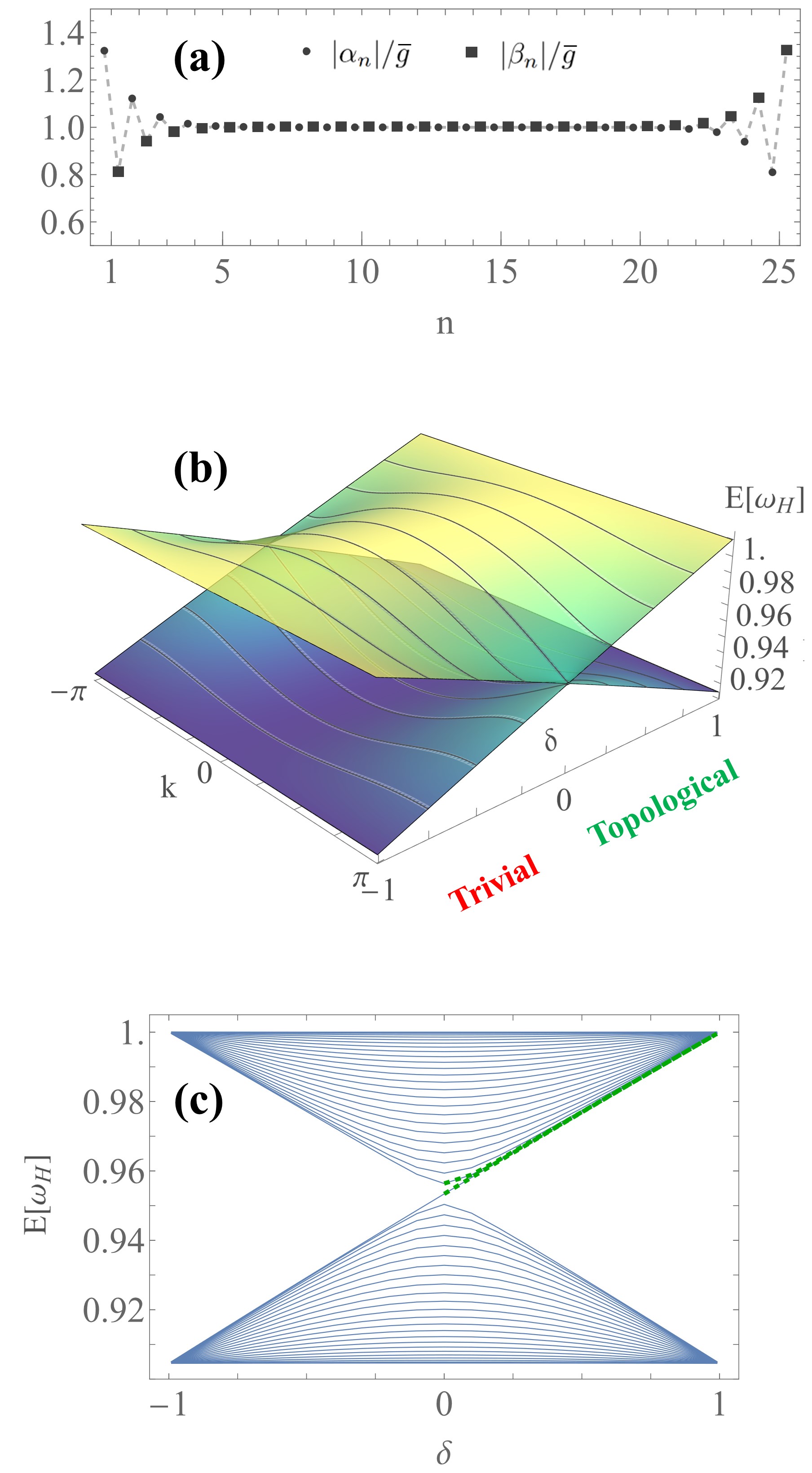} 
  \caption{\textbf{Full-chain semiclassical ground-state and spectrum.} (a) The semiclassical solutions $|\alpha_n|$ and $|\beta_n|$ for a chain satisfying OBC with $\mu=0.9$, $\delta=0.5$ and $N=25$ cells. (b) Band structure as a function of $\delta$ for $\mu=0.1$. (b) Spectrum for open boundary conditions as a function of $\delta$ for $\mu=0.1$ and $N=25$. Each line identify an energy level. Dashed green lines indicate quasi-degenerate levels emerging near the upper band.}
  \vspace*{-3mm}
  \label{f. OBC sol}
\end{figure}

\textit{Full-chain --} A similar symmetry-breaking transition is observed in the extended system, when the inter-cell cross-Kerr terms are present $\epsilon_2\neq 0$. In this case, the homogeneous and inhomogeneous regimes are reminiscent of  the Mott insulating and density wave phases found in the extended Bose-Hubbard model in the limit of vanishingly small tunneling \cite{kuehner1998,rossini2012}. 
To investigate the emergence of the topological phase, we focus on the homogeneous regime $\mu<1$ that features an effective quadratic coupling between the resonators above threshold, as it will be clear in the following. To derive analytical solutions, we will take the limit in which all Kerr and cross-Kerr nonlinearities are small, and the low-energy spectrum is well described by a Gaussian approximation~\cite{felicetti_universal_2020}.
We start by determining the semiclassical equilibrium points which identify the centers of the maxima appearing in the Husimi function. We thus introduce the displaced operators, 
\be c_{A n}=d_{A n}-\alpha_n \quad {\rm and} \quad c_{B n}=d_{B n}-\beta_n, \ee 
in the Hamiltonian $H$ and we choose the complex numbers $\alpha_n$ and $\beta_n$  so that all terms linear in  $d_{A n}$ and $d_{B n}$ vanish and the resulting quadratic Hamiltonian is positive definite.
By doing so one can show that $\alpha_n$ and $\beta_n$   are purely imaginary and satisfy the following equations
\bea
&\!\!\!\!\!\!\!\!2\,|\alpha_n|^2+ \mu\,(1-\delta)\,|\beta_{n}|^2 + \mu\,(1+\delta)\, |\beta_{n-1}|^2 =2g^2\label{eq-alphan}\\
&\!\!\!\!\!\!2\, |\beta_n|^2+ \mu\,(1-\delta)\,|\alpha_{n}|^2+ \mu\,(1+\delta)\,|\alpha_{n+1}|^2 =2g^2.\label{eq-betan}
\eea
The above equations are invariant under $\alpha_n\rightarrow-\alpha_n$  and $\beta_n\rightarrow-\beta_n$ yielding $2^{2N}$ degenerate semiclassical ground-states.
This degeneracy stems from the boson-parity preserving nature of the  Hamiltonian $H$.\\ 
The solution of Eqs.~(\ref{eq-alphan})-(\ref{eq-betan})  depends on the boundary conditions. Namely, for a closed chain, {\sl i.e.} periodic boundary conditions (PBC), we obtain $|\alpha_n|^2=|\beta_n|^2=\bar{g}^2$. Conversely,  for an open chain,   {\sl i.e.} open boundary conditions (OBC),  the amplitudes $\alpha_n$ and $\beta_n$ are larger near the edge sites and  decay to the constant PBC solution in the bulk, as shown in  Fig.~\ref{f. OBC sol}(a). In both cases the ground-state energy scales with the number of resonators and the distance from the second-order phase transition. 
When the system is sufficiently far from the transition, $\bar{g}\gg 1$, the Husimi function of the chain consists of multiple disconnected lobes, analogous to the two-resonator case shown in Fig.~\ref{f.two_mode_Husimi}(b). In this regime, we can describe the dynamics  of Gaussian fluctuations around single semiclassical solutions by a quadratic Hamiltonian of the form
\begin{widetext}
\be H_2=\sum_{n} \bigg\{\sum_I \Big[\Omega_{I n} d^\dag_{I n}d_{I n}+ \frac{\Lambda_{In}}{2}(d_{In} d_{I n}+{\rm H.c.})\Big]
 +\Big[J_{1n} (d_{An}^\dag d_{Bn}-d_{An }d_{Bn})+  J_{2n} (d_{An+1}^\dag d_{Bn}-d_{An+1 }d_{Bn})+{\rm H.c.}\Big]\bigg\} \label{eq-h2}
\ee
\end{widetext}
where the tunneling terms, $J_{1n}$ and $J_{2n}$  are given by $J_{1n} = \epsilon_1| \alpha_n \beta_n| $ and $J_{2n} = \epsilon_2 |\alpha_{n+1} \beta_n| $, while  the renormalized frequencies and driving amplitudes read $\Omega_{An}=\lambda+2\epsilon_L|\alpha_n|^2$ and  $\Lambda_{An}=\lambda-2\epsilon_L|\alpha_n|^2$. Analogous relations hold for $\Omega_{Bn}$ and $\Lambda_{Bn}$ with $\alpha_n$ replaced by $\beta_n$. Note that, in the Hamiltonian $H_2$, the particular choice of the semiclassical ground-state, satisfying ${\rm Im}[\a_n]={\rm Im}[\beta_n]>0$, only affects the phase of the tunneling term and the squeezing direction  and it can be reabsorbed in a redefinition of the $d_{In}$ operators.

Equation~\eqref{eq-h2} shows that the Hamiltonian $H_2$ depends on the boundary conditions. This boundary dependence can be traced back to the non-local nature of the nonlinearities, which yields different semiclassical solutions of Eqs.~(\ref{eq-alphan})-(\ref{eq-betan}) under PBC and OBC, and has important physical consequences. In particular, as we demonstrate below, it results in the breaking of the bulk-boundary correspondence. For PBC the Hamiltonian $H_2$ can be diagonalized by switching to $k$-space and performing a Bogoliubov transformation \cite{sup}. By doing so, up to constant terms, we obtain,
$
H_2=\sum_{k,\eta}{\cal E}^\eta_{k}\,C^\dag_{\eta k}C_{\eta k}
$
where $C^\dag_{\eta k}$ and $C_{\eta k}$ with $\eta\in\{+,-\}$  are creation and annihilation operators of  Bogoliubov excitations, and the energies $\mathcal{E}_k^\pm$ can be cast as
\begin{align}\label{e. Ek}
 \mathcal{E}_k^\pm=\omega_H\,\sqrt{\frac{1}{1+\mu}\left(1\pm\mu\,\sqrt{\frac{1+\delta ^2}{2}+\frac{1-\delta ^2}{2} \cos k}\right)}~,
\end{align}
with $\omega_H=2\,\sqrt{\lambda(\lambda-\omega)}$. The spectrum of Bogoliubov excitations has a gap which vanishes for $\epsilon_2=\epsilon_1$, as shown in Fig.~\ref{f. OBC sol}(b). To investigate the topological properties of the two bands for $\delta\neq0$ we calculate the Zak phase,  
\be
\phi_{\rm Zak}^\eta=i\int_{-\pi}^\pi \frac{dk}{\pi}\;\la G|C_{\eta k}\,\partial_k C^\dag_{\eta k}|G\ra~, 
\ee
here $|G\ra$ denotes the ground-state of $H_2$. We obtain~\cite{sup} $\phi_{\rm Zak}^\pm=\pm 1$ for $\delta>0$ and $\phi_{\rm Zak}^\pm=0$ for $\delta<0$, reminiscent of the SSH model.
However, as we now show, differently from  the SSH model, a non-vanishing \(\phi_{\rm Zak}\) does not correspond to the emergence of protected edge modes in the open chain. This is due to the fact that,  under OBC, the Hamiltonian \(H_2\) has non-homogeneous parameters.
Specifically, as illustrated in Fig.~\ref{f. OBC sol}(c), the OBC spectrum does not feature edge states pinned at the center of the gap, as in the standard SSH model. Instead, it displays two additional levels aligned with the lower edge of the upper band for $\epsilon_2\gg\epsilon_1$. These states are not protected by the gap and spread over the entire lattice. 

The structure of the spectrum can be understood focusing on the limit $\delta = 1$. 
In this case, the Hamiltonian $H$ describes a chain composed of $N-1$ resonator pairs, forming the bulk, and two isolated sites at the edges. 
In the bulk pairs, SSB leads to the appearance of a bosonic ``Higgs-like'' mode, whose energy, $\omega_H$, does not depend on the strength of the nonlinear interactions, according to Eq.~\eqref{e. Ek}, and of a ``Goldstone-like'' mode with energy  
$\omega_H\sqrt{(1-\mu)/(1+\mu)}$, vanishing in the limit $\mu=1$. The edge sites instead have a single mode with energy $\omega_H$. Consequently, at $\delta=1$, the edge mode and the Higgs-like bulk modes are degenerate. In this scenario any finite value of $\epsilon_1$ couples all degenerate modes causing the spreading of the edge modes across the whole chain \cite{sup}.

We refer to the bulk modes as ``Higgs'' and ``Goldstone'' since,  for $\mu=1$, they arise from the breaking of the continuous $O(2)$-symmetry, as displayed in Fig.\ref{f.two_mode_Husimi}(d), and reminiscent of the SSB in the Sigma Model yielding a massive ``Higgs'' boson, with mass independent of the quartic term (at tree level), and a massless Goldstone boson \cite{peskin1995,schwartz2014}.
\begin{figure}
  \hspace*{-4mm}\includegraphics[scale=0.55]{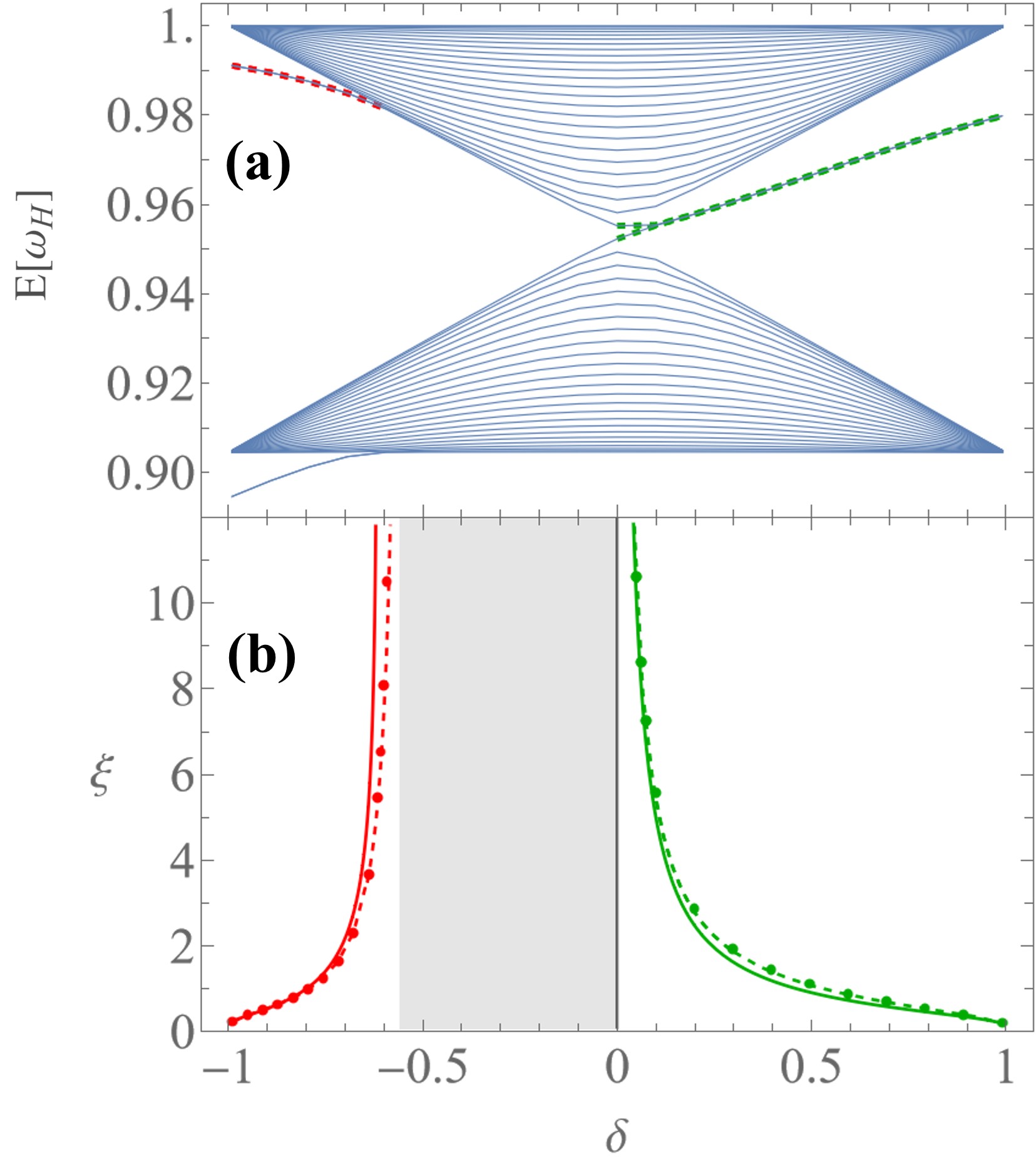} 
    \caption{\textbf{Restoring the bulk-boundary correspondence.} (a) Spectrum for open boundary conditions as a function of $\delta$ for $\mu=0.1$, $N=25$ and $\delta_\lambda=0.02\,\lambda$. The dashed green and red lines indicate topological and impurity-induced boundary modes, respectively. (b) Dependence on $\delta$ of the correlation length $\xi$ in the two cases. Solid lines are analytical fits given, for topological modes, by the SSH-like behavior, and, for impurity-induced, by perturbation theory.}
    \label{f. OBC spectrum drive}
\end{figure}\\

\textit{Edge modes --} The breaking of the bulk-boundary correspondence ultimately stems from the degeneracy of the edge and Higgs-like bulk modes — which, in turn, arises from the non-homogeneous structure of $H_2$ under OBC — it should be possible to restore it by introducing a small non-uniform correction to the quadratic part of the Hamiltonian $H$, thereby making $H_2$ more homogeneous.
This can be achieved by reducing the driving on the left and right boundary sites by adding the following term to $H$:  
\be
 \delta H_\lambda = -\delta_{\lambda} \,\big( c^\dag_{A 1}c^\dag_{A 1} + c^\dag_{B N}c^\dag_{B N} + \text{H.c.} \big),
\ee
with $\delta_\lambda \ll \lambda$.  
The correction $\delta H_\lambda$ modifies the structure of the semiclassical ground state, leading to the emergence of two isolated modes in the spectrum of the quadratic Hamiltonian \cite{sup}, as shown in Fig.~\ref{f. OBC spectrum drive}(a) for $\delta_\lambda=0.02\,\lambda$. These modes appear both around $\delta = 1$ and $\delta = -1$. In the topological phase ($\delta > 0$),  the two modes remain isolated from the rest of the spectrum. In contrast, in the topologically trivial phase ($\delta < 0$), they stay isolated only as long as $\epsilon_2 \lesssim \delta_\lambda$. This highlights the different effect of the correction $ \delta H_\lambda$ in the two phases.
To understand this point it is useful to consider the behavior of the semiclassical ground-state of $H$ under OBC close to $\delta=-1$ and $\delta=1$.
In the first case, it is homogeneous and a finite $\delta_\lambda$ introduces an impurity, giving rise to non-topological localized modes.
On the other hand, as explained above, at $\delta = 1$ the ground-state is inhomogeneous, therefore introducing  $\delta H_\lambda$ reduces the inhomogeneity and favors the emergence of the topological modes.
The different origins of the localization in the two phases is also captured by the behavior of the localization length $\xi$, as shown in Fig.~\ref{f. OBC spectrum drive}(b). In the topological phase, $\xi$ follows an SSH-like behavior, $\xi^{-1} \propto \log{(\epsilon_2/\epsilon_1)}$, whereas in the trivial phase, standard perturbation theory yields a correlation length which diverges when $\delta$ reaches a threshold value \cite{sup}.\\

In conclusion, we considered a nonlinear many-body bosonic chain where parametric pumping induces a quantum phase transition. We showed that, in the symmetry-broken phase, the presence of staggered cross-Kerr terms gives rise to emergent bosonic modes with a topological band structure. Although in general this nonlinearity-driven topological phase breaks the bulk-boundary correspondence, we identified boundary conditions for which topological edge modes are restored. This phenomenology could be observed with current solid-state quantum technologies~\cite{beaulieu_observation_2023,busnaina_quantum_2024,slim_optomechanical_2024}, and it could be directly applied in quantum sensing applications~\cite{mcdonald_exponentially-enhanced_2020,bao_exponentially_2022,arandes_quantum_2024,sarkar_critical_2024,beaulieu_criticality-enhanced_2024}. This work points at spontaneous symmetry breaking as a mechanism to generate topological phases in nonlinear quantum systems.\\

{\it Acknowledgments --} We acknowledge useful discussions with B. Van Heck. A.C., S.F. and V.B. acknowledge financial support from
PNRR MUR project PE0000023-NQSTI financed by the
European Union – Next Generation EU. S.F. acknowledges support from CQSense project financed by Fondazione Compagnia di San Paolo.

\bibliography{BiblioF}

\end{document}


\preprint{APS/123-QED}

\title{Supplementary material for \\
``Non-linearity driven topology {\sl via} spontaneous symmetry breaking"  }
\author{Alessandro Coppo\textsuperscript{\textcolor{blue}{*}}\hspace*{-0.5mm}}
\affiliation{Istituto dei Sistemi Complessi, Consiglio Nazionale delle Ricerche, Via dei Taurini, 19 I-00185 Roma (IT)}
\affiliation{Physics Department,  University of Rome, ``La Sapienza'', P.le A. Moro, 2 I-00185 Roma (IT)}
\author{\vspace*{4mm}Alexandre Le Boit\'e}
\affiliation{Universit\'e Paris Cit\'e, CNRS, Mat\'eriaux et Ph\'enom\`enes Quantiques, F-75013 Paris, France}
\author{Simone Felicetti}%
\affiliation{Istituto dei Sistemi Complessi, Consiglio Nazionale delle Ricerche, Via dei Taurini, 19 I-00185 Roma (IT)}
\affiliation{Physics Department,  University of Rome, ``La Sapienza'', P.le A. Moro, 2 I-00185 Roma (IT)}
\author{Valentina Brosco}%
\affiliation{Istituto dei Sistemi Complessi, Consiglio Nazionale delle Ricerche, Via dei Taurini, 19 I-00185 Roma (IT)}
\affiliation{Physics Department,  University of Rome, ``La Sapienza'', P.le A. Moro, 2 I-00185 Roma (IT)}


\maketitle

\begin{center}
\vspace*{-25mm}\textcolor{blue}{$^*$}(\href{mailto:alessandro.coppo@cnr.it}{\footnotesize alessandro.coppo@cnr.it})\end{center}

\vspace*{10mm}
\section{The periodic chain}

\noindent \textbf{\textit{Semiclassical ground-state}}\\
Let us consider the chain of $2N$ nearest-neighbor Kerr resonators introduced in the main text Eqs. (1)-(2). For $\lambda\gg\omega$ and $\mu<1$, the amplitudes $\alpha_n,\beta_n$ identifying the semiclassical ground-states are purely imaginary and satisfy Eqs.~(6)-(7) of the main text. It is useful to define the parameter $\tau>0$ via
\begin{equation}\label{e. tau}
    \sinh^2{\left(\frac{1}{2\tau}\right)}=\frac{1/\mu^2-1}{1-\delta^2}~,
\end{equation}
and rewrite the equations as
\begin{align}
    &\dfrac{1}{4\sinh^2{\left(\frac{1}{2\tau}\right)}}\left(|\alpha_{n+1}|^2+|\alpha_{n-1}|^2-2\,|\alpha_{n}|^2\right)-|\alpha_{n}|^2+\bar{g}^2=0\qquad\! \mbox{for} \qquad n=2,...,N\label{e. semiclass eq a}\\
    &|\beta_{n}|^2=g^2-\mu\left[\frac{1-\delta}{2}\,|\alpha_{n}|^2+\frac{1+\delta}{2}\,|\alpha_{n+1}|^2\right]\qquad\qquad\qquad\qquad\!\mbox{for} \qquad n=1,...,N\label{e. semiclass eq b}
\end{align}
In general, Eq.~\eqref{e. semiclass eq a} is solved by
\begin{equation}\label{e. an}
    |\alpha_{n}|^2=\bar{g}^2+c_1\, e^{-n/\tau}+c_2\, e^{n/\tau}~,
\end{equation}
with the constants $c_1,c_2$ fixed by the boundary conditions. The parameter $\tau$ thus takes the meaning of a correlation length which describes the semiclassical ground-state inhomogeneity along the chain.
In the case of Periodic Boundary conditions (PBC) 
\begin{equation}\label{e. PBC a}
    \begin{dcases}
      \alpha_{N+1}=\alpha_1~,\\
      \alpha_N=\alpha_0~,
    \end{dcases}
\end{equation}
Eqs.~\eqref{e. PBC a} lead to $c_1=c_2=0$ yielding the homogeneous solution
\begin{equation}\label{e. PBC sol}
    |\alpha_{n}|^2= |\beta_{n}|^2=\bar{g}^2~.
\end{equation}
Notice that the second line of Eqs.~\eqref{e. PBC a} is equivalent to $\beta_N=\beta_0$ via Eq.~\eqref{e. semiclass eq b}.\\
\\
\noindent \textbf{\textit{Quadratic Hamiltonian}}\\
According to Eq.~(8) of the main text, in the case of PBC and for $\mu<1$, the quadratic Hamiltonian describing Gaussian fluctuations around a single semiclassical ground-state can be cast as
\begin{equation}\label{e. PBC H2}
    H_2= \sum_{n=1}^N \bigg\{\sum_{I\in\{A,B\}} \Big[\,\Omega\, d^\dag_{I n}d_{I n}+ \frac{\Lambda}{2}\left( d_{In} d_{I n}+{\rm H.c.}\right)\Big]
 +\Big[J_{1} (d_{An}^\dag d_{Bn}-d_{An }d_{Bn})+  J_{2} (d_{A,n+1}^\dag d_{Bn}-d_{A,n+1 }d_{Bn})+{\rm H.c.}\Big]\bigg\} 
\end{equation} 
with
\begin{equation}\label{e. PBC parameters}
\begin{dcases}
    \Omega=\lambda+\frac{\lambda-\omega}{1+\mu}~,\quad \Lambda=\lambda-\frac{\lambda-\omega}{1+\mu}~,\\
    J_1=\frac{\mu(\lambda-\omega)}{1+\mu}\,\frac{1-\delta}{2}~,\quad J_2=\frac{\mu(\lambda-\omega)}{1+\mu}\,\frac{1+\delta}{2}~.
\end{dcases}
\end{equation}
\vspace*{2mm} \\ 
Given the PBC $d_{AN+1}=d_{A1}$, $d_{BN}=d_{B0}$, and, being all the parameters in \eqref{e. PBC H2} homogeneous, it is natural to switch to the $k$-space via the discrete Fourier transform. We thus define the ladder operators $d_{Ik}=N^{-1/2}\sum_n d_{In}\, e^{-ikn}$, with $k=2s\pi/N$ restricted to the first Brillouin zone (FBZ), \textit{i.e.} $s\in[-N/2+1,\,N/2]$ for $N$ even or $s\in[-(N-1)/2,\,(N-1)/2]$ for $N$ odd, so that Eq.~\eqref{e. PBC H2} reads

\begin{equation}\label{e. PBC H2 k-space}
    H_2= \sum_{k\in\mathrm{FBZ}} \bigg\{\sum_{I\in\{A,B\}} \Big[\,\Omega\, d^\dag_{I k}d_{Ik}+ \frac{\Lambda}{2}\left( d_{Ik}d_{I,-k}+{\rm H.c.}\right)\Big]
 +\Big[J_{k}\, (d_{Ak}^\dag d_{Bk}-d_{Ak }^\dag d_{B,-k})+{\rm H.c.}\Big]\bigg\} 
\end{equation} 
with
\begin{equation}\label{J crossK}
  J_k=\frac{\mu(\lambda-\omega)}{1+\mu}\left(\frac{1-\delta}{2}+\frac{1+\delta}{2}\,e^{-ik}\right)~.
\end{equation}
Introducing the normal modes operators, $C_{\eta k}$ and $C_{\eta k}^\dagger$, defined as 
\begin{equation}\label{e. def C modes}
C_{\eta k}=\frac{1}{\sqrt{2}}\left(Z_{\eta Ak} +\eta\,\frac{J_k}{|J_k|} \, Z_{\eta Bk}\right)~,
    \end{equation}
with $\eta\in\{+,-\}$ and the operators
\begin{equation}\label{e. def Z}
Z_{\eta Ik}=\cosh(\nu_k^\eta)\,d_{Ik}+\sinh(\nu_k^\eta)\,d_{I,-k}^\dagger~,\qquad \tanh\left(2\nu_k^\eta\right)=\frac{\Lambda-\eta\,|J_k|}{\Omega+\eta\,|J_k|}~,
\end{equation}
the Hamiltonian \eqref{e. PBC H2 k-space} is diagonalized as $H_2=\sum_{k,\eta}\mathcal{E}_k^\eta\, C_{\eta k}^\dagger C_{\eta k}$, up to constant terms.  
The energy spectrum is therefore described by two bands defined as
\begin{align}\label{e. Ek Sup}
 \mathcal{E}_k^\pm=\omega_H\,\sqrt{\frac{1}{1+\mu}\left(1\pm\mu\,\sqrt{\frac{1+\delta ^2}{2}+\frac{1-\delta ^2}{2} \cos k}\right)}~,
\end{align}
with $\omega_H=2\sqrt{\lambda(\lambda-\omega)}$. 
Each $\mathcal{E}_k^\pm$ identifies a one-particle doubly degenerate excited level due to the $k\rightarrow -k$ symmetry. Notice the levels $k=0,\pi$ have no degeneracy. The total number of levels is then reduced from $2N$ to $N+2$ for $N$ even, or to $N+1$ for $N$ odd.\\
From Eq.~\eqref{e. Ek Sup} one can easily show that the Gaussian Hamiltonian is positive definite, indeed  $\mathcal{E}_{k}^\pm>0$  for $\forall k\in \mathrm{FBZ}$ and $\mu<1$. This indicates that the semiclassical solution \eqref{e. PBC sol} is stable in the regime $\mu<1$.\\
\\
\noindent \textbf{\textit{Topological invariant: Zak phase}}\\
For one-dimensional systems a widely employed topological invariant is the Zak phase defined by
\begin{equation}
\phi_{\rm Zak}^\eta=i\int_{-\pi}^\pi \frac{dk}{\pi}\;\braket{ G|C_{\eta k}\,\partial_k C^\dag_{\eta k}|G}~.
\end{equation}
The definition of the operators $Z_{\eta I k}$ given in \eqref{e. def Z} has only real coefficients, therefore we can  use \eqref{e. def C modes} and relate the Zak phase with the argument of $J_k=|J_k|e^{-i\theta_k}$, {\sl i.e.}
    \begin{equation}
      \phi_{\mathrm{Zak}}^\eta=\frac{\eta}{2\pi}\int_{-\pi}^{\pi}dk\,\partial_k \theta_k=\frac{\eta}{2\pi i}\oint_{J_k}\frac{dw}{w}=\eta\,{\rm Ind}_{J_k}(0)~,
    \end{equation}
where ${\rm Ind}_{J_k}(0)$ is the winding number around the origin of the curve $J_k:(-\pi,\pi]\rightarrow\mathbb{C}$ and represents the total number of times that the curve travels counterclockwise around that point. Using Eq.~\eqref{J crossK}, we thus obtain 
\begin{equation}\label{e. Zak phase for SSH}
    \begin{cases}
       \delta>0\rightarrow \phi_{\mathrm{Zak}}^\eta=\eta\neq 0~,\\
       \delta<0 \rightarrow\phi_{\mathrm{Zak}}^\eta=0~.
    \end{cases}
\end{equation}
This implies that in the SSB regime $\lambda\gg\omega$ for $\mu<1$, the chain enters a topological non-trivial phase when $\delta>0$ ($\epsilon_2>\epsilon_1$). At the topological phase transition the band gap  vanishes according to Eq.~\eqref{e. Ek Sup}.
\\

\section{The open chain}

\begin{figure}[H]
\centering
  \hspace*{-5mm}\includegraphics[scale=0.7]{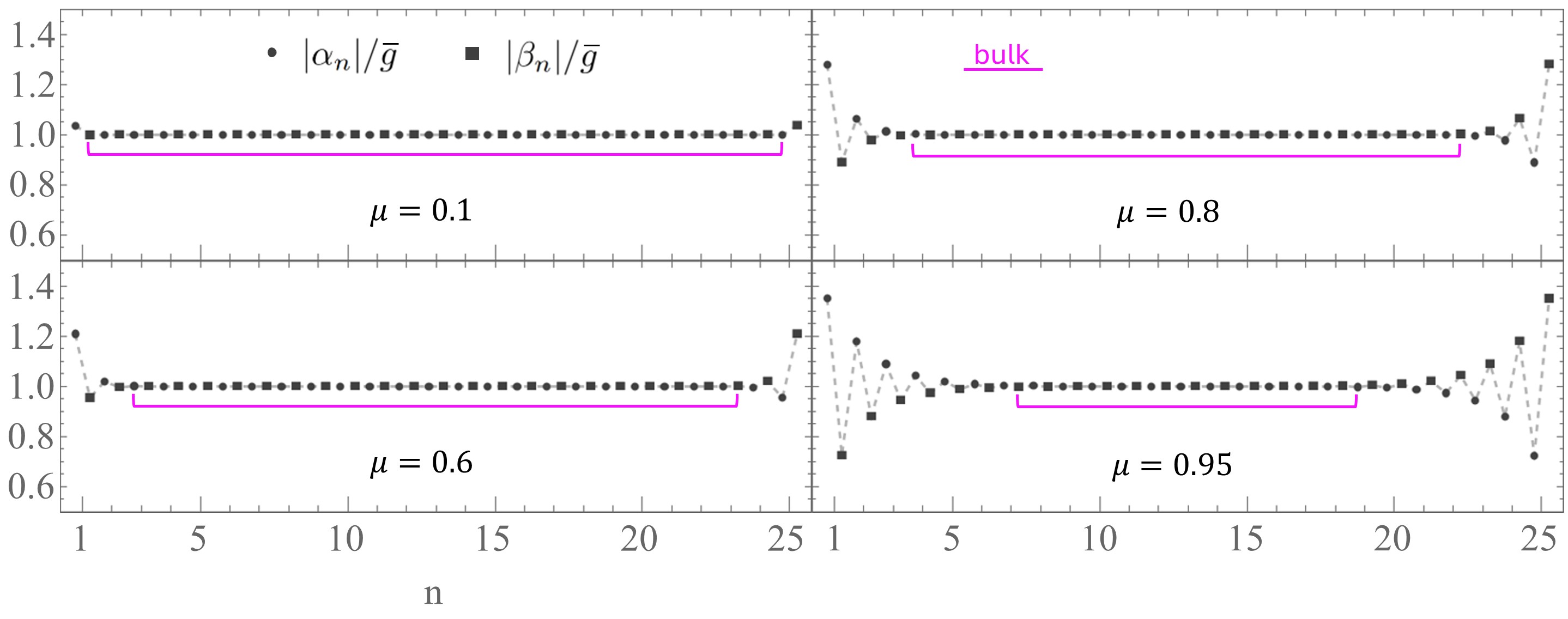}
   \caption{The OBC semiclassical ground-states for a chain of $N=25$ cells with $\delta=0.5$ and different values of $\mu$. The magenta lines identifies the bulk regions, \textit{i.e.} the cells satisfying $10\,\tau<n<N+1-10\,\tau$. }
    \label{f.S1}
\end{figure}

\begin{figure}[H]
    \centering
  \begin{minipage}[c]{0.55\textwidth}
    \includegraphics[scale=0.38]{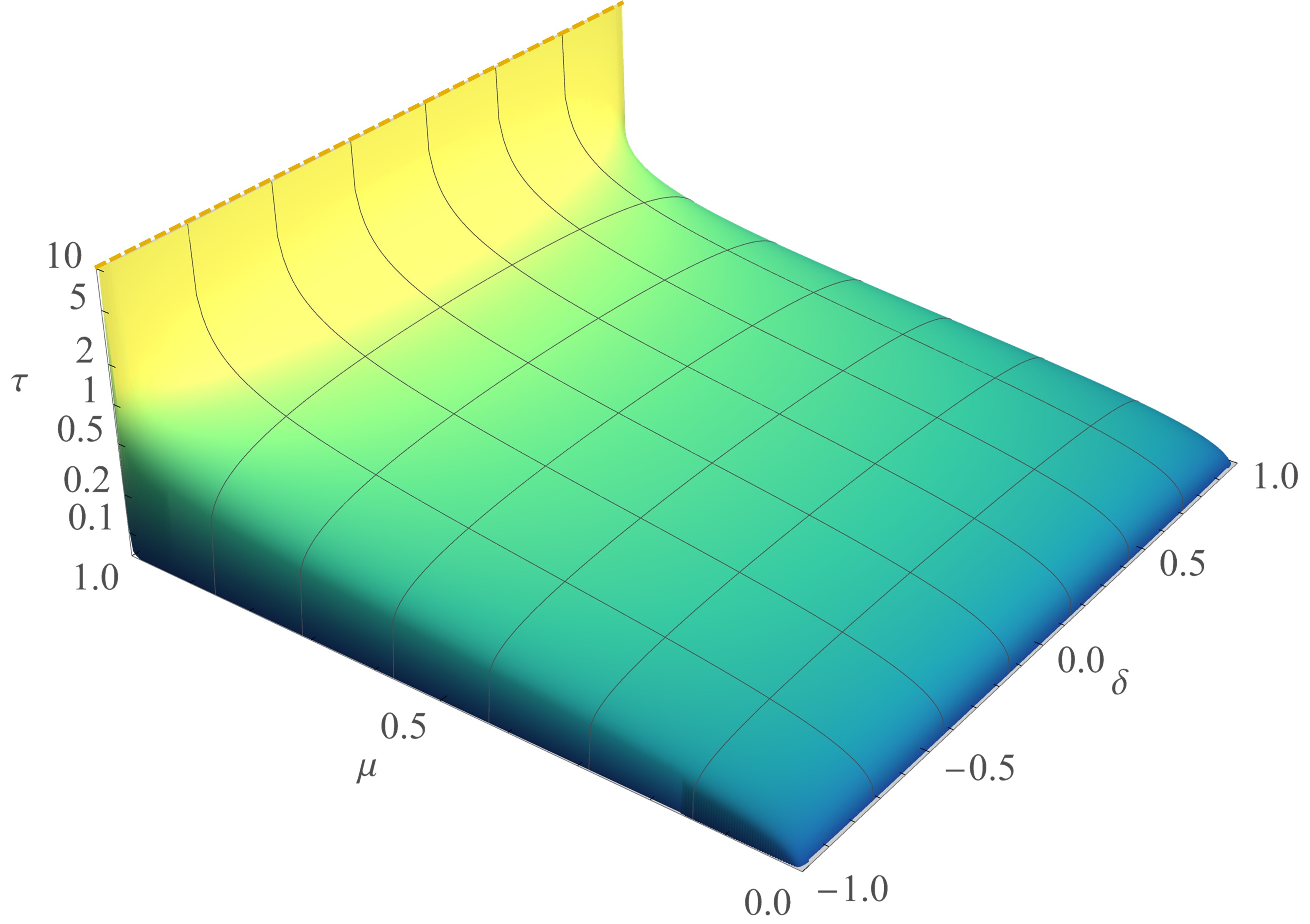} 
  \end{minipage}
  \begin{minipage}[c]{0.4\textwidth}
  \caption{The correlation length $\tau$ 
  as a function of $\mu$ and $\delta$ (log-scale).}
  \label{f.S2}
  \end{minipage}
\end{figure}

\noindent \textbf{\textit{Semiclassical ground-state}}\\
Let us now consider the chain in case of Open Boundary Conditions (OBC), \textit{i.e.} $\alpha_{N+1}=\beta_0=0$. The semiclassical ground-states are still given by Eq.~\eqref{e. an}, but with the conditions
\begin{equation}\label{e. OBC a}
    \alpha_{N+1}=0~,\qquad
    \left[1-\mu^2\left(\frac{1-\delta^2}{2}\right)^2\right]|\alpha_1|^2-\mu^2\,\frac{1-\delta^2}{4}|\alpha_2|^2=g^2\left(1-\mu\,\frac{1-\delta}{2}\right)~,
\end{equation}
with the second relation equivalent to $\beta_0=0$ via Eq.~\eqref{e. semiclass eq b}. Inserting Eqs.~\eqref{e. OBC a} in Eqs.~\eqref{e. an} and \eqref{e. semiclass eq b} for $\mu<1$, we find the solution
\begin{align}\label{e. OBC sol}
    &|\alpha_{n}|^2= \bar{g}^2\left[\frac{4}{R}\cdot\sinh{\left(\frac{N+1-n}{\tau}\right)}+1-e^{-\left(\frac{N+1-n}{\tau}\right)}\right]~,\qquad|\beta_{n}|^2=|\alpha_{N+1-n}|^2~,
\end{align}
with $R=\mu\,e^{N/\tau}\left[(1-\delta)\,e^{1/\tau}+(1+\delta)\right]$. 
Notice that, as shown by Fig.~\ref{f.S1}, Eq.~\eqref{e. OBC sol} tends to  Eq.~\eqref{e. PBC sol} for $n\sim N/2$. In particular the correlation length $\tau$, whose behavior is plotted in Fig.~\ref{f.S2}, defines a "bulk region", $10\,\tau\,${\small $\lesssim$} $n$ {\small $\lesssim$} $N+1-10\, \tau$, where $|\alpha_n|^2\sim|\beta_n|^2\sim\bar{g}^2$. When $\mu$ approaches $1$, the length $\tau$ exponentially increases and the size of the bulk region is reduced. On the other hand, if $\mu\ll 1$, $\tau$ tends to zero and all the chain, except the first and last sites, can be considered as the bulk.\\

\vspace*{2mm}
\begin{figure}[H]
    \centering
  \begin{minipage}[c]{0.55\textwidth}
    \includegraphics[scale=0.65]{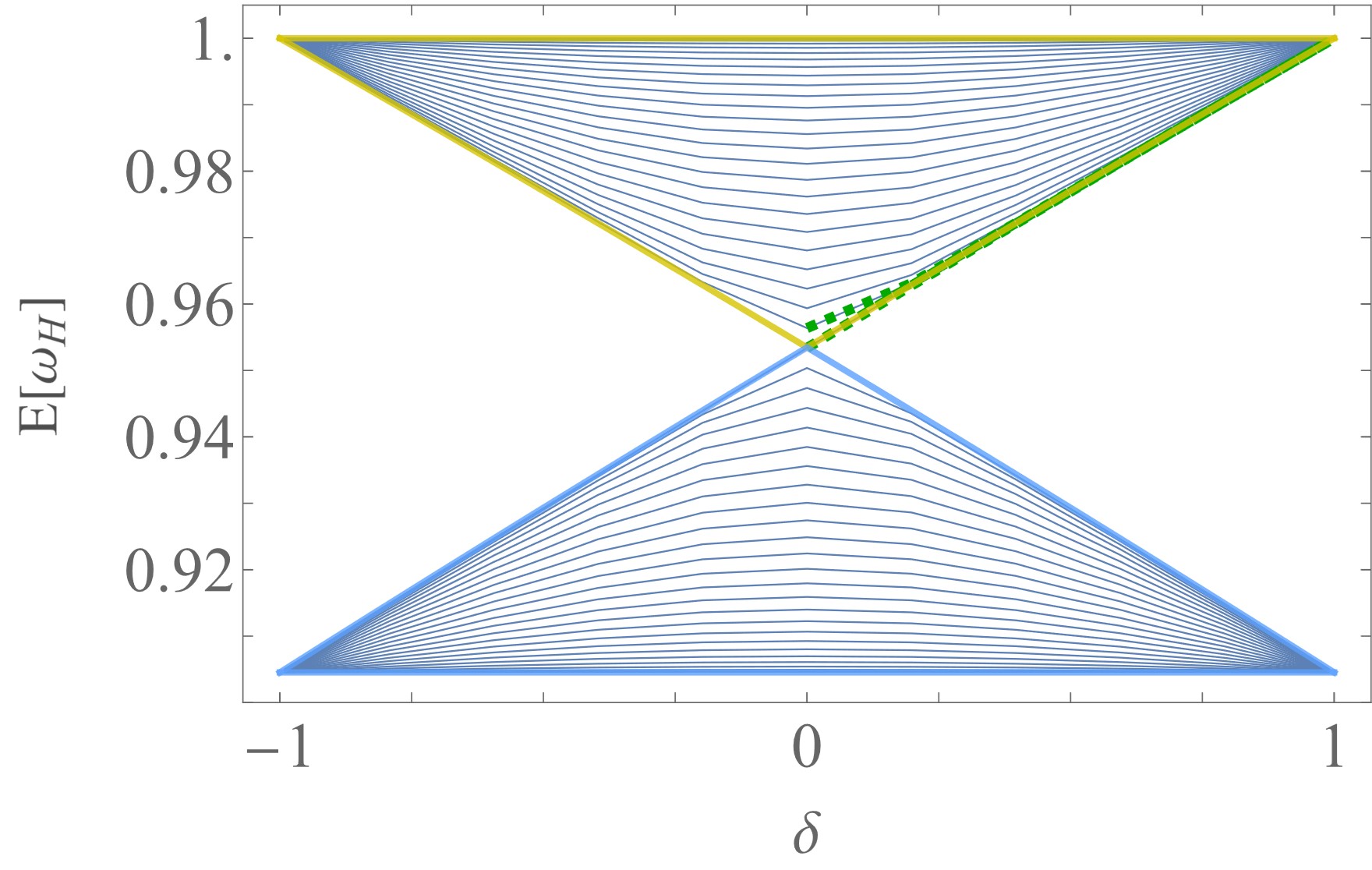} 
  \end{minipage}
  \begin{minipage}[c]{0.4\textwidth}
  \caption{The OBC energy spectrum as a function of $\delta$, for $\mu=0.1$ and $N=25$. Each line identifies a level $\mathcal{E}_\nu$. The yellow and cyan thick lines represent the edges of the upper and lower band, respectively. Dashed green lines indicate quasi-degenerate levels emerging near the lower edge of the upper band. }
  \label{f.S3}
  \end{minipage}
\end{figure}

\begin{figure}[H]
    \centering
  \begin{minipage}[c]{0.55\textwidth}
    \includegraphics[scale=0.65]{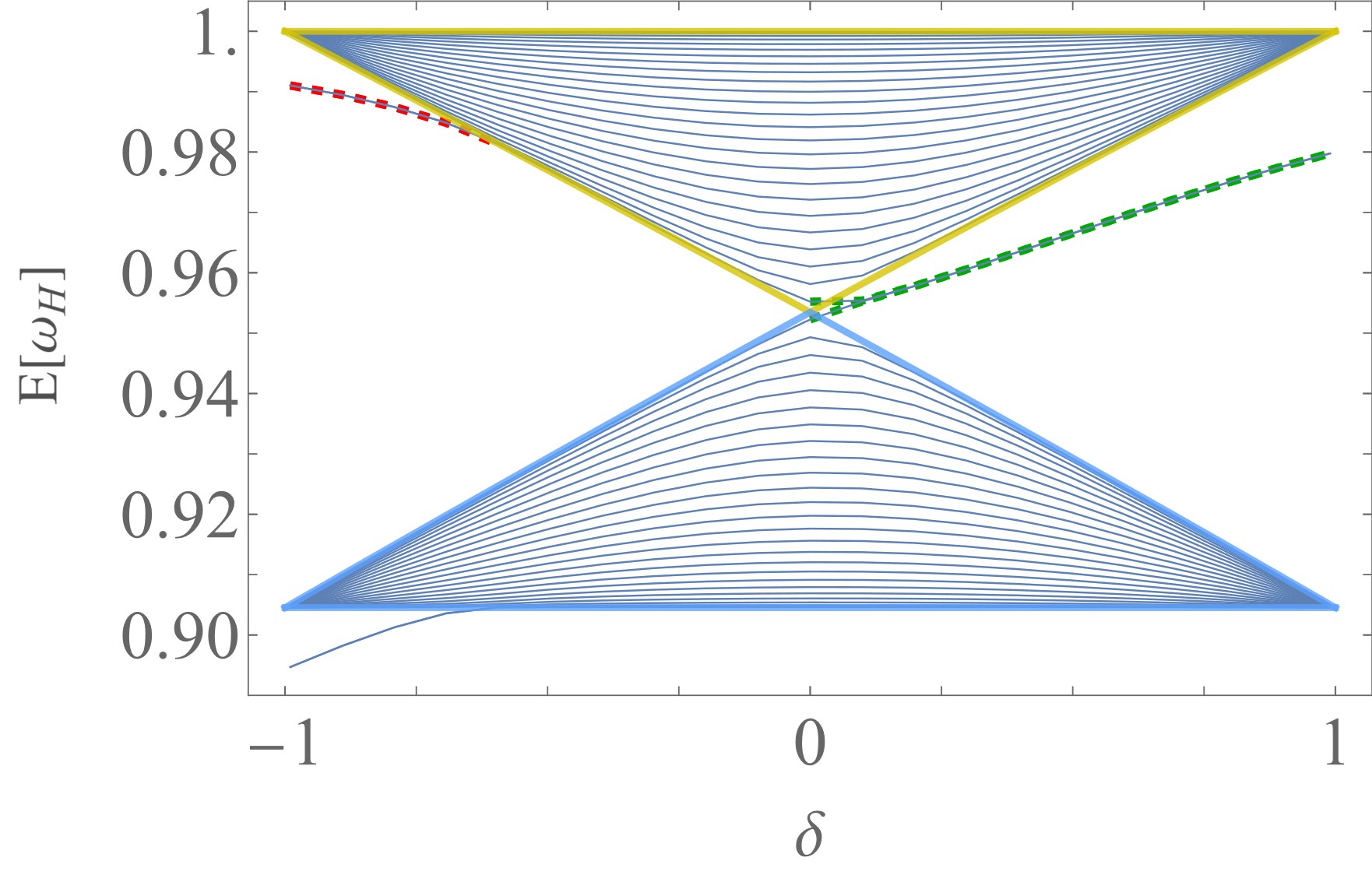} 
  \end{minipage}
  \begin{minipage}[c]{0.4\textwidth}
  \caption{The OBC energy spectrum as function of $\delta$, for $\mu=0.1$, $N=25$ and $\delta_\lambda=0.02\,\lambda$. Colors as in Fig.~\ref{f.S3}. Dashed lines indicate energy levels lying inside the band gap and describing localized states.}
  \label{f.S4}
  \end{minipage}
\end{figure}

\begin{figure}[H]
  \begin{minipage}[c]{0.52\textwidth}
    \includegraphics[scale=0.5]{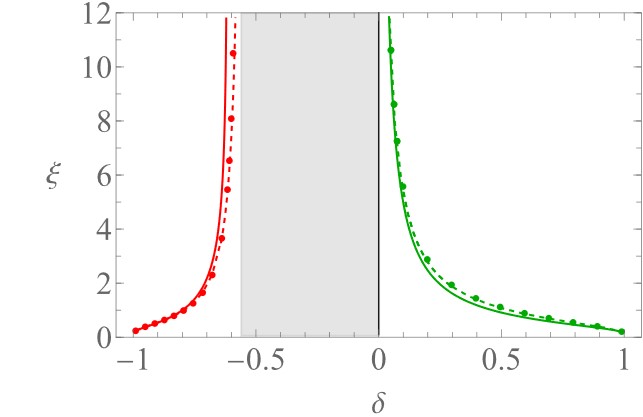} 
  \end{minipage}
  \begin{minipage}[c]{0.4\textwidth}
  \caption{The correlation length $\xi$ as a function of $\delta$ for topological (green) and spurious (red) boundary modes with $\mu=0.1$, $N=25$, $\delta_\lambda=0.02\,\lambda$. The dots are obtained via numerical exponential fits of the Hamiltonian eigenstates and the dashed lines combining Eq.~\eqref{e. xi def} with numerical eigenvalues. The solid lines are analytical estimations given by the ``SSH behavior" in the topological phase, or by perturbation theory in the trivial phase, as in Eq.~\eqref{e. xi imp}.}
  \label{f.S8}
  \end{minipage}
\end{figure}

\noindent \textbf{\textit{Quadratic Hamiltonian}}\\
It is useful to rewrite the quadratic Hamiltonian given by Eq.~(8) of the main text as
\begin{equation}\label{e. Bdg H}
H_2=\sum_{In,I'n'}\left[X_{In,I'n'}\,d_{In}^\dagger d_{I'n'}+\frac{1}{2}\left(Y_{In,I'n'}\,d_{In}^\dagger d_{I'n'}^\dagger+Y_{In,I'n'}^*\,d_{In} d_{I'n'}\right)\right]~,
\end{equation}
with the matrix elements $X_{In,I'n'}$ and $Y_{In,I'n'}$ related to the semiclassical ground-state  given in Eq.~\eqref{e. OBC sol}.
In this way $H_2$ can be diagonalized \textit{ \`a la} Bogolubov introducing the $4N\times4N$ matrix 
\begin{equation}
    \mathcal{L}=
    \begin{pmatrix}
          X & Y \\
          -Y^* & -X^*
     \end{pmatrix}
\end{equation}
and numerically calculating its spectrum
$\mathcal{L}\,\vec{u}_\nu=\mathcal{E}_\nu\vec{u}_\nu$, with $\vec{u}_\nu=\left(u_{In\nu},\widetilde{u}_{In\nu}\right)$, $\sum_{In}(\,|u_{In\nu}|^2-|\widetilde{u}_{In\nu}|^2\,)=1$.
The positive eigenvalues $\mathcal{E}_\nu$ identify the energy levels of $H_2$, that can be therefore expressed as  $H_2=\sum_\nu\mathcal{E}_\nu C_\nu^\dagger C_\nu+\mathcal{E}_0$, with $C_\nu=\sum_{In}(u_{In\nu}d_{In}-\widetilde{u}_{In\nu}d_{In}^\dagger)$ and $\mathcal{E}_0$ a constant term.
The emerging energy spectrum is plotted in Fig.~\ref{f.S3} where the edges of the  upper and lower band,  given by Eq.~\eqref{e. Ek Sup}, are highlighted in color.\\
\noindent The eigenstates $\vec{u}_\nu$ solve the following  equations: 
\begin{equation}\label{e. OBC states}
    \begin{dcases}
        \Omega_{An}\, u_{An}+\Lambda_{An}\,\widetilde{u}_{An}+J_{1n}\left(u_{Bn}-\widetilde{u}_{Bn}\right)+J_{2n}\left(u_{Bn-1}-\widetilde{u}_{Bn-1}\right)=\mathcal{E}\,u_{An}~,\\
        \Omega_{Bn}\,u_{Bn}+\Lambda_{Bn}\,\widetilde{u}_{Bn}+J_{1n}\left(u_{An}-\widetilde{u}_{An}\right)+J_{2n}\left(u_{An+1}-\widetilde{u}_{An+1}\right)=\mathcal{E}\,u_{Bn}~,\\
        \Omega_{An}\, \widetilde{u}_{An}+\Lambda_{An}\,u_{An}-J_{1n}\left(u_{Bn}-\widetilde{u}_{Bn}\right)-J_{2n}\left(u_{Bn-1}-\widetilde{u}_{Bn-1}\right)=-\mathcal{E}\,\widetilde{u}_{An}~,\\
        \Omega_{Bn}\, \widetilde{u}_{Bn}+\Lambda_{Bn}\,u_{Bn}-J_{1n}\left(u_{An}-\widetilde{u}_{An}\right)-J_{2n}\left(u_{An+1}-\widetilde{u}_{An+1}\right)=-\mathcal{E}\,\widetilde{u}_{Bn}~,
    \end{dcases}
\end{equation}
where we removed the index $\nu$ for the sake of a lighter notation and we used the site-dependent parameters $\Omega_{In}$, $\Lambda_{In}$ and $J_{In}$ introduced in the main text, which in the case of PBC reduce to Eqs.~\eqref{e. PBC parameters}.
In the bulk region Eq.~\eqref{e. OBC states} can be unraveled to obtain
\begin{equation}\label{e. eigenstates}
\frac{1}{4\cosh^2{\left(\frac{1}{2\xi}\right)}}\left(u_{An+1}+u_{An-1}-2u_{An}\right)+u_{An}=0~,\qquad\widetilde{u}_{An}=\frac{\mathcal{E}-\lambda}{\mathcal{E}+\lambda}~u_{An}
\end{equation}
with
\begin{equation}\label{e. xi def}
\cosh^2{\left(\frac{1}{2\xi}\right)}=\frac{1-\chi^2}{1-\delta^2}~,\qquad \chi=\frac{1}{\mu}-\left(\frac{1}{\mu}+1\right)\,\frac{\mathcal{E}^2}{\omega_H^2}~.
\end{equation}
Analogous equations hold for $u_{Bn}$, $\widetilde{u}_{Bn}$. Eq.~\eqref{e. eigenstates} is solved by
\begin{equation}\label{e. energy eigenstate}
    u_{An}=(-1)^n\,\left(w_1\,e^{-n/\xi}+w_2\,e^{n/\xi}\right)~,
\end{equation}
where the constants $w_1$, $w_2$ can be fixed imposing Eq.~\eqref{e. OBC states} right outside the bulk region.
The solution \eqref{e. energy eigenstate} describes both localized and extended modes. Specifically,  the eigenstates $\vec{u}_\nu$ are exponentially localized  if $\mathrm{Im}(\xi)=0$, or spread across the chain if $\mathrm{Im}(\xi)\neq 0$. According to Eq.~\eqref{e. xi def}, it can be shown that 
\begin{equation}\label{e. loc condition}
    \mathrm{Im}(\xi)=0\;\;\Longleftrightarrow\;\;\mathcal{E}_{k=\pi}^-<\mathcal{E}<\mathcal{E}_{k=\pi}^+~,
\end{equation}
namely, an eigenstate is localized if and only if the related energy lies inside the band gap. In this case the localization length is given by Eq.~\eqref{e. xi def}. As shown by Fig.~\ref{f.S3}, the OBC energy spectrum does not feature isolated levels inside the band gap. Therefore, even when $\delta>0$, the system does not possess localized edge modes.\\
\\
\noindent \textbf{\textit{Death and revival of edge states via an effective model}}\\
As described in the main text, the disappearance of edges states in the topological phase can be understood by considering the spectrum at $\delta = 1$. In this case, the chain is composed of $N-1$ resonator pairs, forming the bulk, and of two isolated sites at the edges. Each bulk pair contains two modes, the Higgs-like mode with energy $\omega_H$ and the Goldstone-like mode with energy $\omega_H\,\sqrt{(1-\mu)/(1+\mu)}$. The edge sites have instead a single mode with energy $\omega_H$ so that, at $\delta=1$, the edge modes and the Higgs-like bulk modes are degenerate. In this scenario any finite value of $\epsilon_1$ ($\delta\lesssim 1$) couples all degenerate modes causing the spreading of the edge modes across the whole chain.
To illustrate this point we consider the following model: 
\begin{equation}\label{e. eff SSH}
H_2^{eff}=\omega_H\sum_{l=1}^{N+1}\,\zeta_l^\dagger\zeta_l+t\sum_{l=1}^{N}\left(\zeta_l^\dagger\,\zeta_{l+1}+\zeta_l\,\zeta_{l+1}^\dagger\right)\quad\mbox{with}\quad\begin{dcases}
    \zeta_1,\zeta_{N+1}\rightarrow\mbox{edge creation operators}~,\\
    \zeta_{l=2,...,N}\rightarrow\mbox{bulk creation operators}~.
\end{dcases}
\end{equation}
Writing Eq.~\eqref{e. eff SSH} in the form \eqref{e. Bdg H}, the matrix $Y$ vanishes and $X$ is a Toepliz tridiagonal matrix which can be exactly diagonalized so that $H_2^{eff}=\sum_\nu E_\nu \gamma_\nu^\dagger\gamma_\nu$, with
\begin{equation}\label{e. Toepliz spectrum}
    E_\nu=\omega_H+2t\cos\left(\frac{\pi\nu}{N+2}\right)\quad\mbox{and}\quad \gamma_\nu=\sqrt{\frac{2}{N+2}}\,\sum_{l=1}^{N+1}\sin\left(\frac{\pi\nu}{N+2}\right)\zeta_l~,\quad \nu=1,...,N+1~.
\end{equation}
All $\gamma_\nu$ modes are thus delocalized across the chain. This behaviour is ultimately rooted in the degeneracy of all the modes for $t=0$. A drastic change happens when introducing a small correction $\delta_{\omega_H}$ to the edge energies via
\begin{equation}
\delta H_2^{eff}=-\delta_{\omega_H}\left(\zeta_1^\dagger\,\zeta_1+\zeta_{N+1}^\dagger\,\zeta_{N+1}\right)~.
\end{equation}
In this case, using second order perturbation theory for $t\ll\delta_{\omega_H}$, the energy of the edge modes becomes
\begin{equation}\label{e. Eedge perturb}
    E_e=\omega_H-\delta_{\omega_H}-\frac{t^2}{\delta_{\omega_H}}+o(t^3)~.
\end{equation}
Solving the eigenvalues equation for $H_2^{eff}+\delta H_2^{eff}$, it is easy to show that, being $E_e<\omega_H$, the corresponding mode is exponentially localized.\\
\\
\noindent \textbf{\textit{Lowering the driving at the boundaries}}\\
The energy $\omega_H$ of the edge modes at $\delta = 1$ depends on the driving intensity $\lambda$. It can be thus reduced by appropriately decreasing the driving at the chain boundaries by an amount $\delta_\lambda$. This reduction is achieved by introducing a correction $\delta H_\lambda$ to the Hamiltonian (see Eq.~(11) of the main text). In this way, for $\lambda \gg \omega$, we obtain a non-vanishing $\delta_{\omega_H}$ leading to the emergence of two isolated modes in the spectrum of the quadratic Hamiltonian, as shown in Fig.~\ref{f.S4}. These two modes appear both in the topological and in the trivial phase, around $\delta = 1$ and $\delta = -1$, and lie in the band gap. They are thus localized according to Eq.~\eqref{e. xi def}. \\ 
\noindent However, in the two phases, the localization arises from different mechanisms.
\begin{itemize}
\item{\textit{Topological phase}}\\
\vspace*{-3mm}\\
When $\delta=1$ ($\epsilon_1=0$), the semiclassical ground state is inhomogeneous for $\delta_\lambda=0$. In particular the length \eqref{e. tau} tends to zero and the inhomogeneity is confined to the first and last sites, \textit{i.e.} $|\alpha_1|=|\beta_N|=g>\bar{g}=|\alpha_{n\neq 1}|=|\beta_{n\neq N}|$ as in Fig.~\ref{f.S5}. According to Eq.~(8) of the main text, the energy of edge and bulk modes can be obtained diagonalizing the following Hamiltonians:%
\begin{align}
    & H_{\mathrm{edge}}^{(\delta=1)}=\Omega_e\,d^\dagger d+\frac{\Lambda_e}{2}\,(d d+d^\dagger d^\dagger)\qquad\mbox{with}\qquad
    \begin{dcases}
    \Omega_e=\lambda+2\epsilon_Lg^2~,\\
    \Lambda_e=\lambda-2\epsilon_Lg^2~,
    \end{dcases}
\end{align}
for the edge sites $d\in\{d_{A1},d_{BN}\}$, and\\ 
\\
\begin{minipage}[c]{0.63\textwidth}
\begin{align}
     & H_{\mathrm{bulk}}^{(\delta=1)}=\Omega\,(d_{A}^\dagger d_{A}+d_{B}^\dagger d_{B})+\frac{\Lambda}{2}\,(d_{A} d_{A}+d_{B}d_{B}+H.c.)\nonumber\\
    &\qquad\;\;+J_2\,(d_A^\dagger d_B-d_A d_B+H.c.)\nonumber
    \end{align}
\end{minipage}
\begin{minipage}[c]{0.4\textwidth}
\begin{align}
    \mbox{with}\qquad\begin{dcases}
\Omega=\lambda+2\epsilon_L\bar{g}^2~,\\
    \Lambda=\lambda-2\epsilon_L\bar{g}^2~,\\
    J_2=\epsilon_2\,\bar{g}^2~,\\
    \end{dcases}
\end{align} 
\end{minipage}\\
\vspace*{2mm}\\
for the bulk pairs $(d_B,d_A)\in\{(d_{Bn},d_{An+1})\}_{n=1,...,N-1}$.
Introducing normal modes through Bogoliubov transformations, each bulk pair contains a mode with energy $\omega_H$ identical to that of the edge modes. The specific inhomogeneous structure of the semiclassical solution in the bulk and in the edges, quantified by the ratio $g/\bar{g}$, leads to degenerate bulk and edges modes in the Gaussian Hamiltonian.
When the driving at the boundaries is decreased,  the inhomogeneity is reduced (see Fig.~\ref{f.S5}), via 
\begin{equation}
    |\alpha_1|=|\beta_N|=g\,\sqrt{1-\frac{\delta_\lambda}{\lambda-\omega}}~,
\end{equation}
\vspace*{-13mm}\\
\begin{figure}[H]
    \hspace*{12mm}
  \begin{minipage}[c]{0.47\textwidth}
    \includegraphics[scale=0.65]{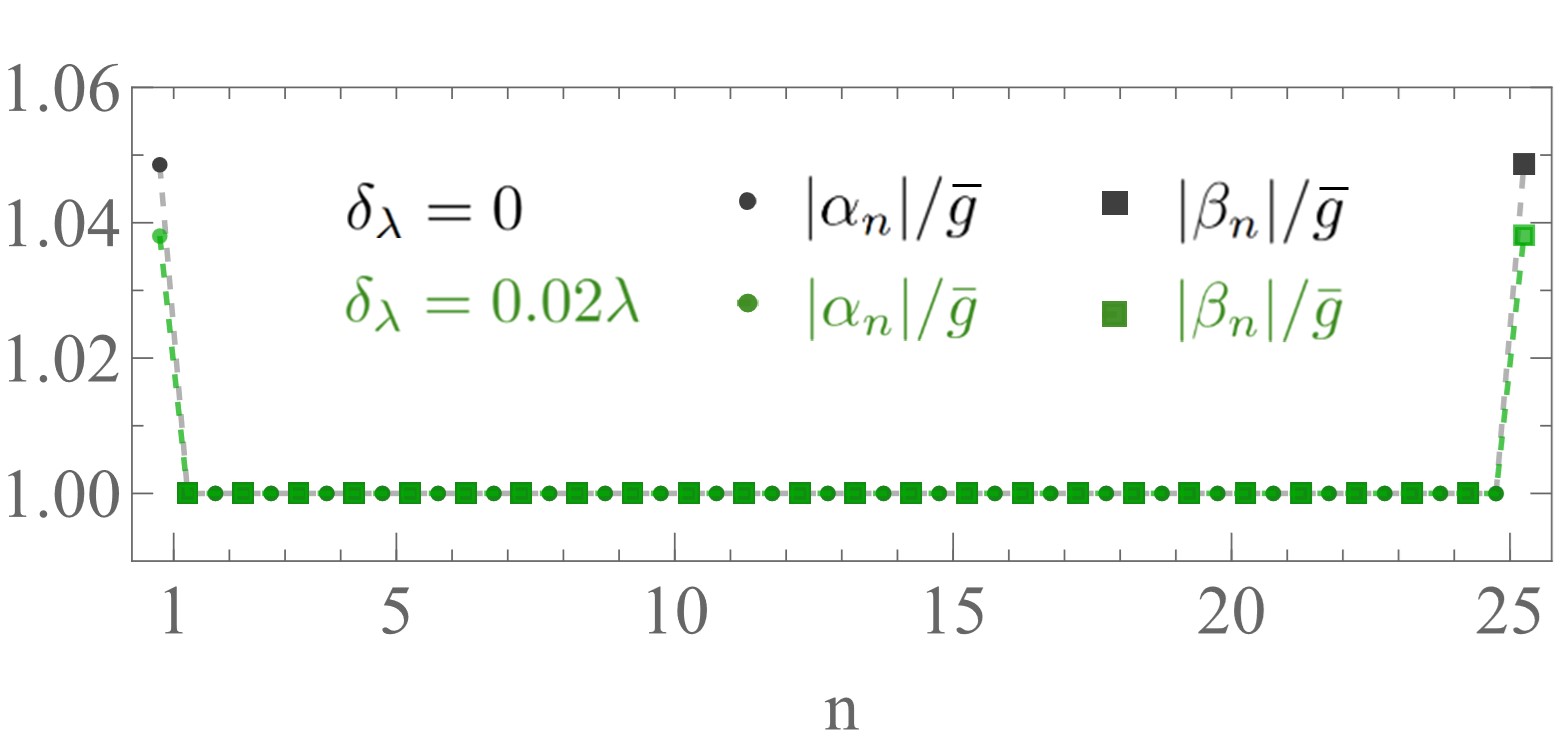} 
  \end{minipage}
  \begin{minipage}[c]{0.4\textwidth}
  \caption{The OBC semiclassical ground-states for $N=25$, $\delta=1$ and $\delta_\lambda=0$ (black) or $\delta_\lambda=0.02\,\lambda$ (green).}
  \label{f.S5}
  \end{minipage}
\end{figure}
and, as in Fig.~\ref{f.S4}, the energy of the edge modes becomes
\begin{equation}
\omega_e^{(\delta=1)}=\omega_H\,\sqrt{\left(1-\frac{\delta_\lambda}{\lambda}\right)\left(1-\frac{\delta_\lambda}{\lambda-\omega}\right)}<\omega_H~.
\end{equation}
In this scenario, introducing a small finite value of $\epsilon_1$ does not couple edge and bulk modes. The former are thus topologically protected modes, localized according to Eq.~\eqref{e. xi def}. 
As long as $\delta_\lambda$ is not too small, the localization length stays well-defined in the entire topological phase, $\delta>0$, and exhibits an ``SSH behavior" $\xi^{-1}\sim\log\left(\frac{1+\delta}{1+\delta}\right)$ (see Fig.~\ref{f.S6} and ~\ref{f.S8}).  More precisely, following Eq.~\eqref{e. xi def}, this behavior emerges when $\chi\sim 0\Leftrightarrow\mathcal{E}\sim (1+\mu)^{-1/2}$, meaning that the edge state energy lies in the center of the band gap. Notice that, in the SSH model, $\delta$ represents the imbalance between inter-cell and intra-cell quadratic tunnelings, whereas, in our case, it corresponds to the imbalance of nonlinear cross-Kerr interactions.\\
\begin{figure}[H]
    \hspace*{13mm}
  \begin{minipage}[c]{0.45\textwidth}
    \includegraphics[scale=0.4]{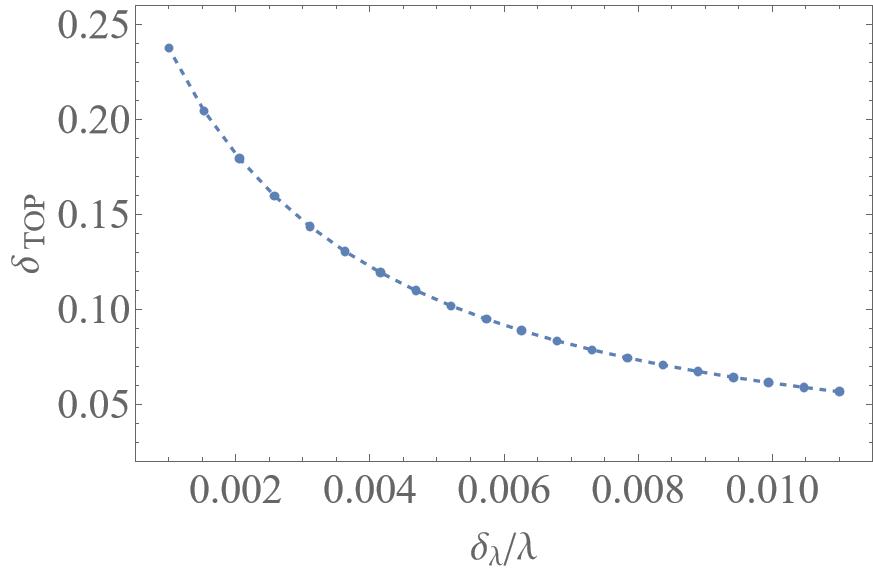} 
  \end{minipage}
  \begin{minipage}[c]{0.4\textwidth}
  \caption{The value $\delta_{\,\mathrm{TOP}}$ below which the localization length of the two modes appearing in the band gap exceeds half the chain length $N/2$ ($\mu=0.1$ and $N=25$). Considering Eq.~\eqref{e. energy eigenstate}, if $\delta>\delta_{\,\mathrm{TOP}}$, the modes can be considered localized.}
  \label{f.S6}
  \end{minipage}
\end{figure}
Let us highlight that $\delta_\lambda$ cannot be nevertheless too large, as this would cause the edge energy at $\delta=1$ to fall outside the band gap, crossing the upper edge of the lower band.\\
Notice that so far we considered the case $\mu\ll 1$. As $\mu$ increases, the bulk region shrinks (see Fig.~\ref{f.S1} right panels) and restoring the edge states requires a reduction of the driving intensities on all sites affected by the inhomogeneity.
\\

\item{\textit{Trivial phase}}\\
\vspace*{-3mm}\\
When $\delta=-1$ ($\epsilon_2=0$), the semiclassical ground-state, for $\delta_\lambda=0$, is homogeneous, \textit{i.e.} $|\alpha_n|=|\beta_n|=\bar{g}$ as in Fig.~\ref{f.S7}. The entire chain, consisting of $N$ decoupled pairs, can thus be considered bulk. According to Eq.~(8) of the main text, the energy of all the pairs can be obtained diagonalizing the following Hamiltonian:\\
\\
\begin{minipage}[c]{0.63\textwidth}
\begin{align}
     & H_{\mathrm{bulk}}^{(\delta=-1)}=\Omega\,(d_{A}^\dagger d_{A}+d_{B}^\dagger d_{B})+\frac{\Lambda}{2}\,(d_{A} d_{A}+d_{B}d_{B}+H.c.)\nonumber\\
    &\qquad\;\;+J_1\,(d_A^\dagger d_B-d_A d_B+H.c.)\nonumber
    \end{align}
\end{minipage}
\begin{minipage}[c]{0.4\textwidth}
\begin{align}
    \mbox{with}\qquad\begin{dcases}
\Omega=\lambda+2\epsilon_L\bar{g}^2~,\\
    \Lambda=\lambda-2\epsilon_L\bar{g}^2~,\\
    J_1=\epsilon_1\,\bar{g}^2~,\\
    \end{dcases}
\end{align} 
\end{minipage}\\
\vspace*{2mm}\\
and $(d_B,d_A)\in\{(d_{Bn},d_{An})\}_{n=1,...,N}$.
The normal modes of all the pairs are thus degenerate and any small finite value of $\epsilon_2$ couples nearest-neighbor pairs spreading all the modes across the whole chain according to Eq.~\eqref{e. Toepliz spectrum}. In this case, decreasing the driving at the boundaries introduces an impurity in the first and last pairs, making the semiclassical ground states inhomogeneous (see Fig.~\ref{f.S7}). The semiclassical solution on the boundary sites can be cast as
\begin{equation}
    |\alpha_1|=|\beta_N|=\bar{g}\,\sqrt{1-\frac{\delta_\lambda}{(1-\mu)(\lambda-\omega)}}~,\qquad
    |\beta_2|=|\alpha_{N-1}|=\bar{g}\,\sqrt{1+\frac{\mu\,\delta_\lambda}{(1-\mu)(\lambda-\omega)}}~.
\end{equation}
\vspace*{-7mm}
\begin{figure}[H]
    \hspace*{12mm}
  \begin{minipage}[c]{0.47\textwidth}
    \includegraphics[scale=0.65]{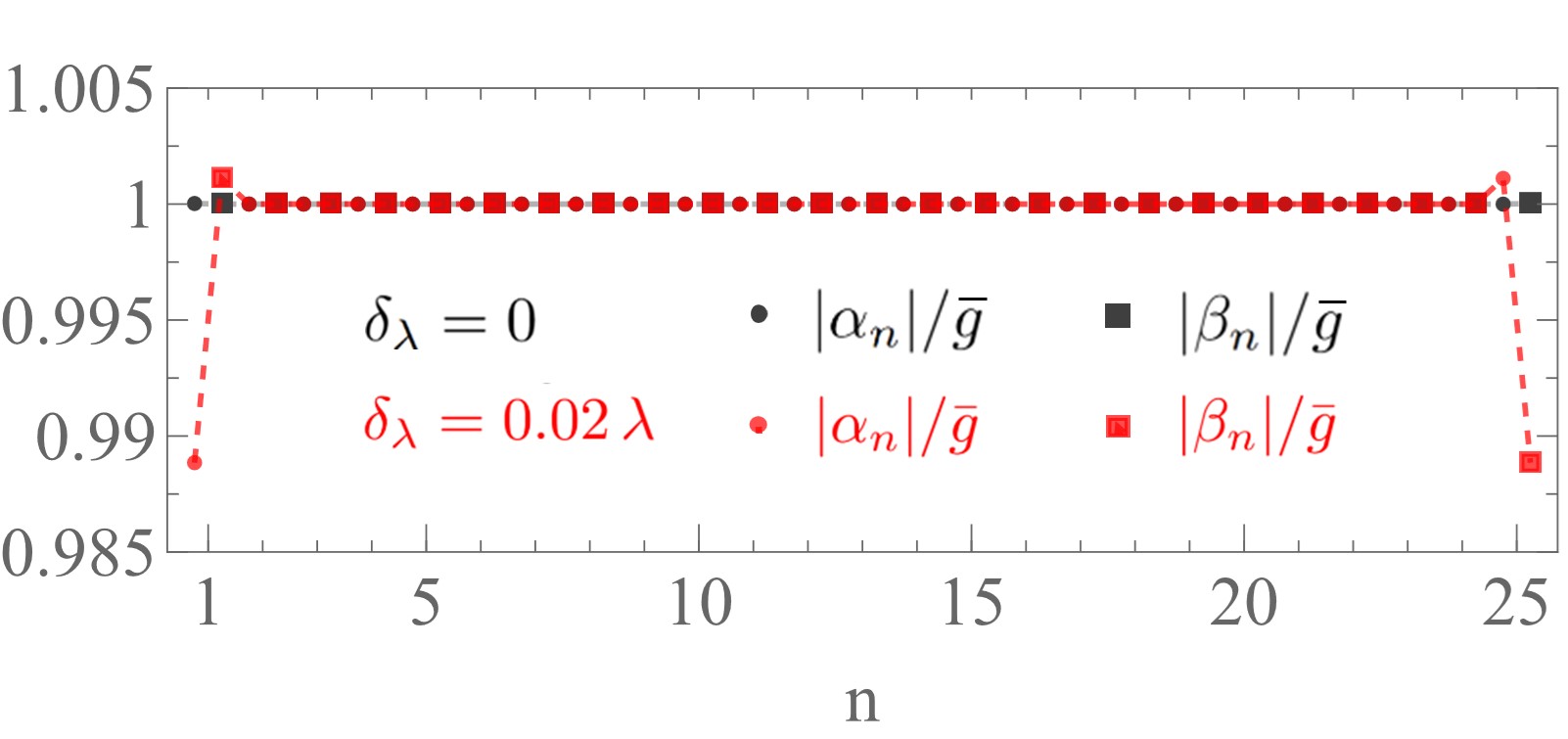} 
  \end{minipage}
  \begin{minipage}[c]{0.4\textwidth}
  \caption{The OBC semiclassical ground-states for $N=25$, $\delta=-1$ and $\delta_\lambda=0$ (black) or $\delta_\lambda=0.02\,\lambda$ (red).}
  \label{f.S7}
  \end{minipage}
\end{figure}
yielding two spurious boundary modes with energy
\begin{equation}
\omega_{\mathrm{spur}}^{(\delta=-1)}=\omega_H\,\left(1-\frac{2\lambda-\omega}{4(\lambda-\omega)}\,\frac{\delta_\lambda}{\lambda}+o\left(\frac{\delta_\lambda}{\lambda}\right)^2\right)<\omega_H~,
\end{equation}
which, as in Fig.~\ref{f.S4}, lie in band gap.
Therefore, also in this scenario, introducing a small finite interaction $\epsilon_2$ between the pairs does not couple boundary and bulk modes. The former are thus localized according to Eq.~\eqref{e. xi def}. Anyway the localization is here induced by the impurity, and it is not related to topology. This different nature is captured by the behavior of the localization length (see Fig.~\ref{f.S8}), which can be estimated via Eqs.~\eqref{e. xi def} and \eqref{e. Eedge perturb}:
\begin{equation}\label{e. xi imp}
    \xi_{\mathrm{spur}}^{-1}\sim 2\,\mathrm{arcosh}\,\sqrt{\dfrac{1-\dfrac{1}{\mu^2}\left[1-(1+\mu)\left(1-\dfrac{\delta_\lambda}{2\lambda}-\dfrac{\lambda}{8\,\delta_\lambda}\left(\dfrac{\mu\,(1+\delta)}{1+\mu}\right)^{\!\!2}\,\right)^{\!\!2}\,\right]^2}{1-\delta^2}}~,
\end{equation}
where we assumed $\omega=0$ $(\lambda\gg\omega)$ for the sake of a lighter notation. The length becomes imaginary when the energy of the spurious modes crosses the lower edge of the upper band. This happens if $J_2\gtrsim\delta_\lambda$ when the interaction becomes high enough to overcome the impurity. Therefore, when $\delta>\delta_{\mathrm{spur}}$ with
\begin{equation}\label{e. delta spur}
    \delta_{\mathrm{spur}}\sim-1 +2\left(\frac{1}{\mu}+1\right)\frac{\delta_\lambda}{\lambda}~,
\end{equation}
as shown in Fig.~\ref{f.S9}, spurious modes lose their localization and spread across the chain. Unlike the edge modes appearing in the topological phase, these boundary modes are thus not localized throughout the entire trivial phase, but instead exhibit a threshold $\delta_{\,\mathrm{spur}}$ beyond which localization is lost.
\end{itemize}
\begin{figure}[H]
  \hspace*{5mm}\begin{minipage}[c]{0.5\textwidth}
    \includegraphics[scale=0.6]{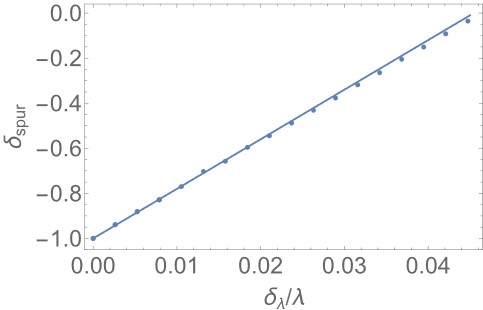} 
  \end{minipage}
  \begin{minipage}[c]{0.4\textwidth}
  \caption{The threshold value $\delta_{\,\mathrm{spur}}$ above which the energy of the boundary spurious modes violates Eq.~\eqref{e. loc condition} causing their spreading across the chain. The solid line is the analytical estimation given by Eq.~\eqref{e. delta spur}.}
  \label{f.S9}
  \end{minipage}
\end{figure}
